# Magnetostatic modes and wave equation

Kirill A. Rivkin

RKMAG Corporation, 651 N. Broad St. Suite 205 462 Middletown Delaware 19709

We propose a mathematical apparatus which converts magnetostatic (Walker) equations into a wave equation by introducing a specific definition of the "magnetic refractive index". Its value can be manipulated by adjusting spatial distribution of the saturation magnetization or the bias magnetic field. We demonstrate that the latter can be accomplished efficiently by utilizing the bias field produced by magnetization patterns recorded onto a neighboring hard magnetic film using the current magnetic storage technology; there is a specific algorithm by which one can convert a two dimensional optical device into its magnetic analogue. The wave equation formalism opens the possibility to adopt a number of optics-like analytical and numerical techniques, including those applicable to scattering, near field, diffraction and geometric (Eikonal equation) optics, as well as a generic finite difference time domain (FDTD) algorithm.

**1. Wave equation.**

It has been long established that in many situations propagation of spin waves is qualitatively similar to that of optical waves, and phenomena such as lensing and diffraction have been experimentally demonstrated[1,2,6]. Physical implementation of such devices often relies on physical alteration of the magnetic media[3,4,5,22], such as thinning out specific sections[8] or physically shaping the waveguides. Other possibilities have been considered as well, such as applying magnetic field within a lens' focal point[7], or locally altering the static, equilibrium magnetization configuration. An attempt has been made to introduce the term "refractive index" to describe behavior of magnetic systems[8] in a context of specific, numerically studied case.

In the present work we intend to propose a general analytical formulation which establishes one to one relationship between the propagation formalism of magnetostatic and optical waves respectively, enabled by a specific definition of "relative magnetic refractive index".

Let us start with a governing equation for magnetostatic modes[9,10] when the magnetization is saturated via a uniform magnetic field $H_0$ applied along the z axis. To describe the relationship between the dynamic magnetization $m_{x,y,z}$ and the dynamic field components $h_{x,y,z}$ we convert a linearized Landau-Lifshitz equation (CGS units) into a matrix form:

$$4\pi \begin{pmatrix} m_x \\ m_y \\ m_z \end{pmatrix} + \begin{pmatrix} h_x \\ h_y \\ h_z \end{pmatrix} = \begin{pmatrix} 1+\chi & -i\kappa & 0 \\ i\kappa & 1+\chi & 0 \\ 0 & 0 & 1 \end{pmatrix} \begin{pmatrix} h_x \\ h_y \\ h_z \end{pmatrix} \qquad (1).$$

Usually in the literature when deriving (Eq.1) the damping contribution is neglected, but in our case we include a Landau-Lifshitz damping term $\frac{\beta}{M_s} M \times (M \times H)$, resulting in the following definitions:

$$\chi = \frac{\omega_0 \omega_M - i\beta \omega_M \omega}{\omega_0^2 - (\omega + i\beta \omega_0)^2} \quad \kappa = \frac{\omega_M(\omega + i\beta \omega_0) - i\beta \omega_0 \omega_M}{\omega_0^2 - (\omega + i\beta \omega_0)^2} \tag{2}$$

$$\omega_M = 4\pi M_s \quad \omega_0 = \gamma H_0 + N M_s + \Delta H$$

Where $M_s$ is the saturation magnetization, $H_0$ is the uniform external field applied to saturate the media, $\Delta H$ is an additional bias field, $\gamma$ is the gyromagnetic ratio, $\omega$ is an operating frequency, N is a demagnetization factor: $4\pi$ if the z axis coincides with the out-of-plane coordinate, zero if its in-plane. We consider the case of a thin magnetic media where $\Delta H$ and $M_s$ are uniform along the thickness but can be non-uniform along the lateral directions. By using a Maxwell's equation $\nabla \cdot \boldsymbol{b} = \nabla \cdot (\boldsymbol{h} + 4\pi \boldsymbol{m}) = 0$ and introducing the magnetostatic potential such that $\boldsymbol{h} = \nabla \varphi$, (Eq.2) becomes:

$$\frac{\partial^2 \varphi}{\partial x^2}(1+\chi) + \frac{\partial^2 \varphi}{\partial y^2}(1+\chi) + \frac{\partial^2 \varphi}{\partial z^2} + \frac{\partial \varphi}{\partial x}\frac{\partial \chi}{\partial x} - i\frac{\partial \varphi}{\partial y}\frac{\partial \kappa}{\partial x} + \frac{\partial \varphi}{\partial y}\frac{\partial \chi}{\partial y} + i\frac{\partial \varphi}{\partial x}\frac{\partial \kappa}{\partial y} = 0 \tag{3}$$

First order derivatives of $\chi$ and $\kappa$ are typically neglected, assuming that the respective physical variables vary instantaneously:

$$(1+\chi)\left[\frac{\partial^2 \varphi}{\partial x^2} + \frac{\partial^2 \varphi}{\partial y^2}\right] + \frac{\partial^2 \varphi}{\partial z^2} = 0 \tag{4}$$

Solutions within areas characterized by different values of $\chi$ are connected by satisfying the boundary conditions. From now on unless otherwise noted we consider the case where the z axis is oriented in the out-of-plane direction, noting the approach remains valid for an arbitrary magnetization orientation. Key observation is that the boundary conditions require the dependence along the film's thickness (z axis) to remain the same irrespective of x and y coordinates. Assuming the magnetic potential's form $\varphi = \varphi(x,y)(a\cos(k_z z) + b\sin(k_z z))$:

$$(1+\chi)\left[\frac{\partial^2 \varphi(x,y)}{\partial x^2} + \frac{\partial^2 \varphi(x,y)}{\partial y^2}\right] - k_z^2 = 0 \tag{5}$$

Until now the derivation mostly followed the textbook examples[9,10]. The novelty being proposed is that for plane wave solutions of the form $\varphi(x,y) = e^{ik_x x + ik_y y}$ one can select a "baseline" case with a corresponding value of $\chi_0$ and a wavevector magnitude $k_0$, i.e.:

$$k_z^2 = -(1+\chi_0)k_0^2 \tag{6},$$

and then for an arbitrary $\chi$ to use (Eq.6) to express $k_z^2$ in (Eq.5), obtaining:

$$\frac{1}{n^2}\left[\frac{\partial^2 \varphi(x,y)}{\partial x^2} + \frac{\partial^2 \varphi(x,y)}{\partial y^2}\right] + k_0^2 \varphi(x,y) = 0 \tag{7},$$

which is identical to Helmholtz wave equation in optics where $n$ is defined as a relative magnetic refractive index:

$$n^2 = \frac{(1+\chi_0)}{(1+\chi)} \tag{8}$$

(Eqs.7-8) are identical in both SI and CGS units, though in the former case $\omega_M = M_s$. Refractive index $n$ depends on wavelength, since both $\chi$ and $\chi_0$ are functions of the operating frequency $\omega$, which relates to the wavevector $k$ as:

$$\omega^2 = (\omega_0 + \gamma D k^2)\left(\omega_0 + \gamma D k^2 + \omega_M \left(1 - \frac{1-e^{-kd}}{kd}\right)\right) \qquad (9),$$

employing a well known solution[9] for a demag field created by a plane wave in a thin film with a thickness $d$. (Eq.9) allows for an alternative definition of the relative refractive index: from $\omega(k, \Delta H) = \omega(k_0, 0)$ one can derive $k = f(k_0, \Delta H) = nk_0$. The end result is equivalent to (Eq.8), except closed analytical expressions are easy to derive only in the limit of small changes in the bias field $\Delta H$ or the saturation magnetization $M_s$.

While the choice of $\chi_0$ and therefore $k_0$ is arbitrary, it is convenient to choose the parameters which correspond to the dominant portion of the magnetic media and zero damping. The latter is so that $k_0$ has zero imaginary part and damping contribution for an arbitrary wavevector $k = nk_0$ is provided by (Eq.8) in a combination with (Eq. 2).

When the saturating field is oriented in-plane along the x axis (z axis remains the out-of-plane direction) the derivation of the wave equation is near identical. Magnetostatic equation becomes:

$$\frac{\partial^2 \varphi}{\partial y^2}(1+\chi) + \frac{\partial^2 \varphi}{\partial x^2} + \frac{\partial^2 \varphi}{\partial z^2}(1+\chi) = 0 \qquad (10).$$

There are two options for the thickness dependence corresponding to the volume and the Damon-Eshbach modes, corresponding to real and imaginary $k_z$ in $\varphi = \varphi(x,y)(a\cos(k_z z) + b\sin(k_z z))$, yielding the expression:

$$k_z^2 = -\left(\frac{k_{0x}^2}{(1+\chi_0)} + k_{0y}^2\right) \qquad (11).$$

For an arbitrary $\chi$ the analogue of (Eq.7) is:

$$\frac{1}{n_y^2}\frac{\partial^2 \varphi(x,y)}{\partial y^2} + \frac{1}{n_x^2}\frac{\partial^2 \varphi(x,y)}{\partial x^2} + k_0^2 \varphi(x,y) = 0 \qquad (12)$$

$$n_y^2 = \frac{\cos^2\theta + \sin^2\theta(1+\chi_0)}{(1+\chi_0)} \qquad n_x^2 = n_y^2(1+\chi) \qquad (13).$$

(Eq.12) is a more general from of (Eq.7), allowing for a birefringent media. Another important difference that in the in-plane case not only the "baseline" media parameters (Eq.13) are being selected via the choice of $\chi_0$, but one also chooses the "baseline" propagation angle $\theta$ with respect to the x axis. The nature of the resulting solution, i.e. the surface or the volume mode, depends on the sign of $(1 + \chi)$ in (Eq.13). For an arbitrary wavevector $k$ and a propagation angle $\acute{\theta}$ the ratio between the wavevectors' amplitudes is determined from (Eqs.12-13) as:

$$\frac{k^2}{k_0^2} = \frac{n_y^2 n_x^2}{(n_x^2 \sin^2\acute{\theta} + n_y^2 \cos^2\acute{\theta})} \qquad (14).$$

In the out-of-plane case $n_x = n_y$.

Assuming a single frequency excitation with $e^{-i\omega t}$ time dependence the wave equation (Eq.12) is mathematically equivalent to:

$$\left[\frac{1}{n_x^2}\frac{\partial^2 \varphi(x,y,t)}{\partial x^2} + \frac{1}{n_y^2}\frac{\partial^2 \varphi(x,y,t)}{\partial y^2}\right] = \frac{1}{c_0^2}\frac{\partial^2 \varphi(x,y,t)}{\partial t^2} \qquad (15),$$

i.e. for $c_o = \omega/k_o$ every solution of (Eq.15) is automatically a solution of (Eq.12) and therefore respectively of either (Eq.4) or (Eq.10). It is often more efficient to apply slow-varying envelope approximation (SVEA) $\dot{\varphi}(x,y,t) = \varphi(x,y,t)e^{-i\omega t}$, converting (Eq.15) into:

$$\left[\frac{1}{n_x^2}\frac{\partial^2 \varphi(x,y,t)}{\partial x^2} + \frac{1}{n_y^2}\frac{\partial^2 \varphi(x,y,t)}{\partial y^2}\right] + k_0^2 \varphi(x,y,t) = -\frac{2ik_0}{c_0}\frac{\partial \varphi(x,y,t)}{\partial t} \quad (16).$$

Steady-state solution $\frac{\partial \varphi(x,y,t)}{\partial t} = 0$ always satisfies either (Eq.4) or (Eq.10) depending on the choice of refractive index components.

## 2. Refractive index.

Considering specifically the out-of-plane case, let us analyze how the refractive index depends on the wavelength (frequency) and the bias field, and how this learning compares to a micromagnetic modeling based on solving a Landau-Lifshitz equation directly. In doing so we assume material parameters typical for Yttrium Iron Garnet[14] (YIG): $4\pi M_s = 0.18$ T, exchange stiffness $A = 3.65 \cdot 10^{-7}$ erg/cm and $D = 5.3 \cdot 10^{-7}$ erg/G·cm, damping parameter $\beta = 0.0004$, media thickness 15nm.

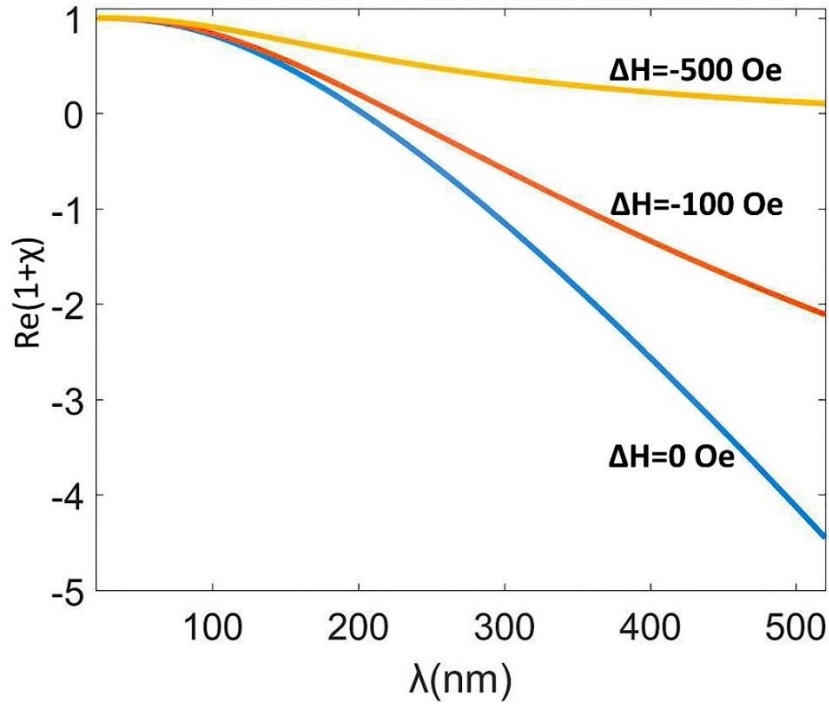

**Figure 1**. Dependence of the real portion of $1 + \chi$ on wavelength.

Consider the wavelength dependence (Fig. 1) of $Re(1 + \chi)$ using (Eq.9). For short wavelengths $Re(1 + \chi) = 0$ when $\omega = \omega_0(\omega_M + \omega_0)$, at which point (Eq.4) has only a trivial solution, propagating waves cannot exist and the value of the magnetic refractive index undergoes discontinuity and becomes imaginary. It is sometimes assumed[10] that below this wavelength, i.e. above the $\omega_0(\omega_M + \omega_0)$ frequency any excitation of magnetostatic modes is impossible, however a more nuanced approach is warranted especially when comparing the results with a micromagnetic modeling, as will be seen below. It can be suggested that when $Re(1+\chi) < 0$, $k_z$ in (Eq.4) becomes imaginary and the surface wave solution with $e^{-z|k_z|}$ and $e^{+z|k_z|}$ dependence is possible, since it satisfies (Eq.4) as well as the boundary conditions at the upper and lower boundaries of the film. Noted, unlike the in-plane case where the surface and the volume waves predominantly propagate in different directions here properties of the volume and the surface mode remain basically the same and therefore the distinction between the two types could be more difficult to observe in an experiment.

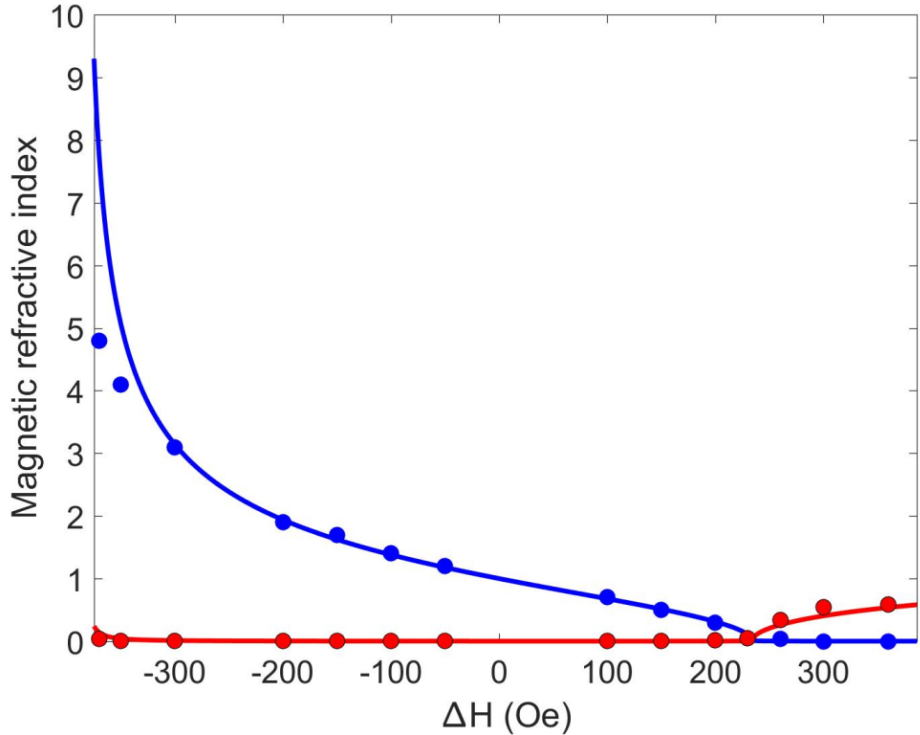

**Figure 2.** Relative magnetic refractive index n as a function of the magnetic bias field ΔH for the fixed frequency $\omega = 1.25 \cdot \omega_0 = 4.05$ GHz and $H_0 = 2900$ Oe. Blue lines represent real, red – imaginary portion, the data depicted by solid lines was obtained from (Eq.8), by dots – via micromagnetic modeling. There is a transition from real to imaginary values occurring around 230 Oe.

In (Fig. 2) we demonstrate the dependence on the bias field $\Delta H$ at the fixed excitation frequency $\omega$. Propagating waves correspond to large real portion of the refractive index and such solutions are available within a specific frequency range, outside of which the value of $n$ becomes imaginary. As seen in (Fig.2) if $\gamma \Delta H > \omega - \omega_0$ in the corresponding portions of the media even the uniform mode's frequency is higher than $\omega_0$, propagation of magnetostatic modes is impossible and instead there is the exponentially decaying solution while the incident wave is reflected back. For a very low $\Delta H$ the value of $\omega_0$ reaches the threshold $\omega_0(\omega_M + \omega_0)$ where the refractive index in (Eq.8) also becomes imaginary. Noted, unlike metals in magnetostatics there are no currents being generated and thus plasmonic phenomena are absent, though boundary-localized ("near-field") modes can be excited.

For a small $|\Delta H|$ a relative refractive index can be approximated as:

$$n \approx 1 - \frac{1}{2}\frac{\gamma \Delta H}{\omega_0}\left(\frac{\omega^2+\omega_0^2}{\omega^2-\omega_0^2}\right) \tag{17}$$

So far every formula prior to (Eq.16) has been derived from magnetostatic (Walker) equations (Eq.4) and (Eq.10) without any additional approximations. However, since a significant portion of modern publications rely on micromagnetic modeling it is worthwhile to consider differences and similarities between micromagnetic results and solutions of the wave equation.

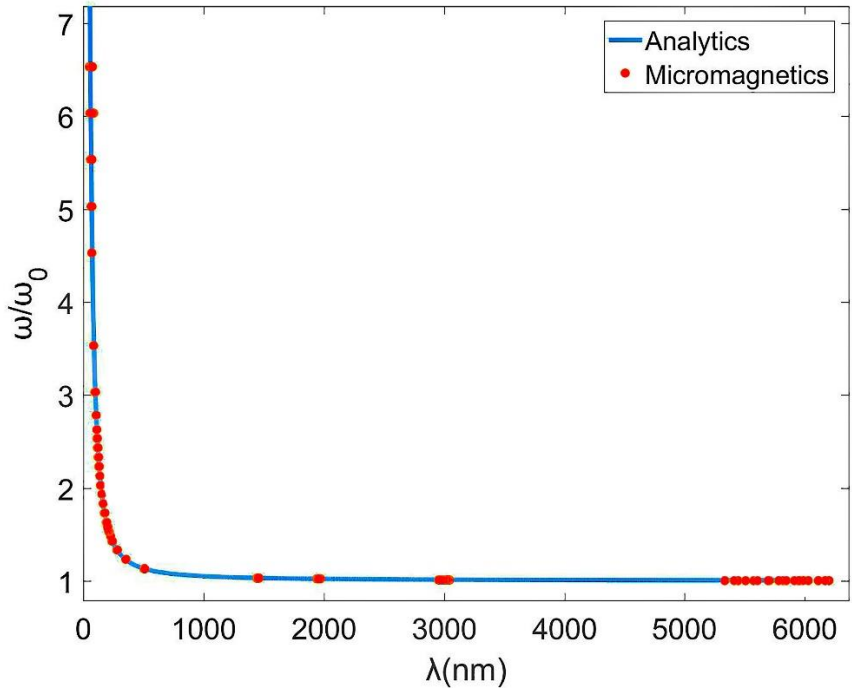

**Figure 3.** Dispersion relationships calculated micromagnetically and analytically by using (Eq.9).

The basic requirement for the two approaches to yield similar results is the ability to reproduce the same dispersion relationship using (Eq.9) and micromagnetically (Fig.3). To do so, the simplest micromagnetic experiment one can propose takes a square sample, discretizes it and applies an in-plane RF field to the central point. Phase of magnetic oscillations in such configuration is expected to increase linearly as $kr$, where $r$ is the distance from the center. Repeating the experiment for different frequencies yields the desired dispersion relationship. Results in (Fig. 3) were obtained via RKMAG micromagnetic software[16] with 15x15x15nm discretization (i.e. less than exchange length) for a 10.5x10.5um media sample with 15nm thickness. Noted, some commercial packages have failed to produce a similar match (Fig.3) and therefore would not be able to reproduce the results of either magnetostatic (Eq.4) or wave (Eq.12) equations. In certain specific cases the divergence is due to a specific choice of boundary conditions[16], which is especially common for the in-plane configuration; then, the (Eq.12) can still be used but with the values of the refractive index values produced by fitting a micromagnetic experiment rather than via analytical formulas.

As long as there is a good match between (Eq.9) and micromagnetics (Fig. 3), reproducing the refractive index dependencies on both the wavelength and the bias field is on the one hand straightforward, but on the other reveals the fundamental difference between the two approaches. To illustrate this point, consider (Fig. 2) where the micromagnetic results were obtained by repeating the point source experiment with different values of the total uniform bias field $H_0 + \Delta H$ and then fitting the refractive index value $n^2 = \frac{k^2(\Delta H)}{k_0^2(\Delta H = 0)}$. Overall there is an excellent match with the magnetostatic theory (Eq.8) and the transition at $\gamma \Delta H = \omega - \omega_0$ exists (Fig.2) in both cases, but at the low $\Delta H$ the results diverge. In a sense, such behavior was already expected given the results in (Fig.3), where micromagnetics has little difficulty exciting spin waves at the wavelengths at and above the critical frequency $\omega = \omega_0(\omega_M + \omega_0)$.

The reason why there is no same discontinuity in micromagnetics as in magnetostatics is that it uses a very different form of thickness dependence: the amplitude of dynamic magnetization is uniform within a discretization cell rather than being described by a cosine function. The two dimensional representation breaks the transition's discontinuous character, but even if one uses extra-fine discretization along the z axis, as long as the thickness is comparable to the exchange length, the actual difference in terms of z dependence between the modes above and below $\omega = \omega_0(\omega_M + \omega_0)$ is small.

This is a common phenomenon related to phase transition behavior in thin films where the surface contribution cannot be ignored and intrinsically three dimensional formulations such as (Eqs.4,10,12) have limitations. Micromagnetic modeling with atomistic-level discretization can in theory produce the most accurate estimation of the behavior in the vicinity of $\omega_0(\omega_M + \omega_0)$ including values of the refractive index in (Fig.2), but it remains prohibitively slow.

Since one of the purposes of the present work is to have a meaningful comparison with the micromagnetics, we choose to employ (Eq.8) to calculate refractive index values except for the points in the vicinity of $\omega = \omega_0(\omega_M + \omega_0)$ where micromagnetics-based fit (Fig. 2) for the two-dimensional discretization lattice are used instead.

There is a practical consideration that for thicker films the discontinuity allows for very high values of the refractive index to be obtainable by either applying the bias field or reducing the saturation magnetization; otherwise the refractive index for short wavelengths remains close to unity since the frequency value in (Eq.9) is completely dominated by the exchange contribution. On the contrary, for large wavelengths (Fig.1) the relative impact of both the bias field and the saturation magnetization increases.

In a related manner when using (Eq.12) with the in-plane geometry one needs to be aware that in micromagnetics with two dimensional discretization the transition between the volume and the Daemon-Eshbach modes, for example, as a function of the propagation angle, is also subdued. When matching micromagnetic and magnetostatic (Eq.12) results one of the options is to use numerically fitted values for $n_x$ and $n_y$, similar to (Fig.2).

There is a common argument that disagreements between magnetostatic and micromagnetic results are impacted by the former "not including the exchange interaction". This is not true since the dynamic fields entering (Eq.1) can have an arbitrary nature, exchange included, and the presence of exchange interaction is explicitly accounted for in (Eq.9).

## 3. Diffraction.

What is the most practical way to use (Eqs.7-17) in order to implement a particular optical device by mimicking a corresponding spatial distribution of the refractive index?

One of the most flexible approaches relies on creating complex bias fields by using a bilayer system where one layer is made from a thin soft magnetic material with low damping ("propagation layer", hereinafter modeled as 15nm thick YIG film), separated by a non-magnetic spacing layer from a hard film ("bias layer") where some magnetization pattern is recorded. The bias layer can include an array of nanodots[11] or a bit pattern media, but a typical granular magnetic recording media is also a good candidate. In the present article we consider the case where the bias layer is a continuous 15nm thick film with the parameters typical for the magnetic recording industry: $4\pi M_s = 1.2$ T (1 to 1.6T values have been used in various products), average grain center to center distance is 7nm with 15% standard variation. Overall the magnetic distributions are such that 45 by 15nm single polarity pattern can be recorded with 1.3nm transition sigma. Effective anisotropy field exceeds 10KOe and therefore hard layer's resonant frequencies are much larger compared to operating frequencies (1-6 GHz) of the propagation layer underneath. It means that while the bias field affects the spin wave propagation, no precession is generated in the hard layer and its typical high damping ($\beta = 0.03$) does not impact the dynamic solutions, which is easily confirmed micromagnetically. Magnetic grains are physically decoupled to reduce the intergranular exchange and conductivity.

Thickness of non-magnetic spacing layer between the bias and the propagation layers is an important optimization parameter. Values below 20nm allow for large (well in excess of 500 Oe) bias fields which are highly localized and have substantial in-plane component, at the expense of a significant local variation of the bias field values due to the granular nature of the media. Above 100nm of separation, the geometry of individual grains has limited impact and the peak value of the out-of-plane component can be made much larger compared to the peak in-plane field components. Noted, a film which is uniformly magnetized in either direction produces zero bias field, so the peak field amplitude in the propagation layer has a non-monotonic dependence on the characteristic size of the magnetization pattern recorded in the bias layer above.

The simplest "spin optics" device is based on diffraction phenomenon. Its functionality can be understood without the wave equation based formalism and the spin diffraction phenomenon has been repeatedly demonstrated experimentally[19,20,21]. Because of the said simplicity diffraction problems allow for fully analytical solutions of (Eq.12) and can be very effective in demonstrating the relative performance of micromagnetic modeling versus numerically solving the wave equation (Eqs.15-16).

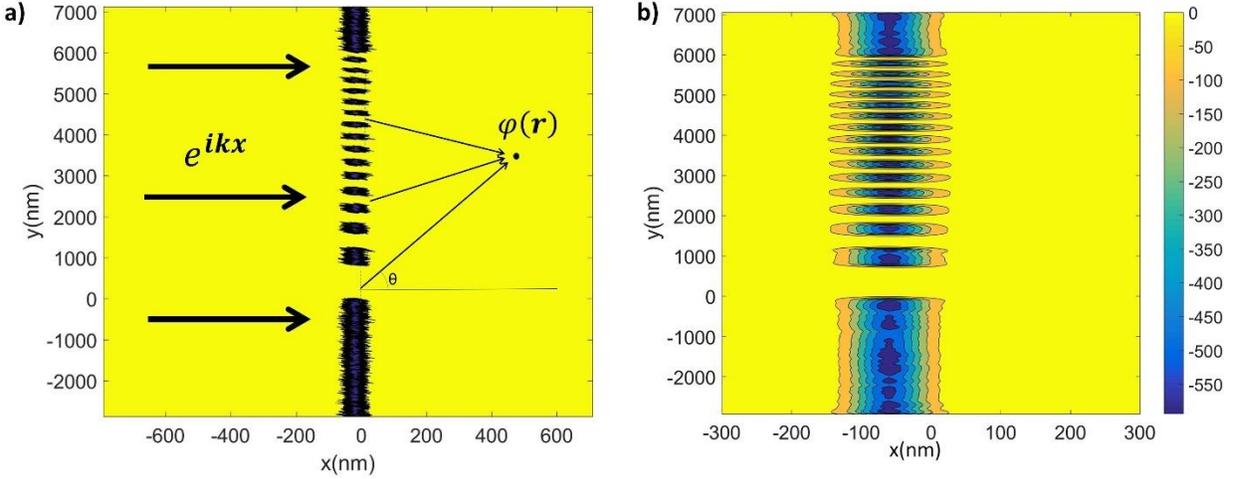

**Figure 4**. **(a)** Diffraction grating recorded onto the granular hard layer, blue color depicts the grains with the magnetization oriented in the opposite direction with respect to the uniform setting field $H_0$. Schematics depicts a spin wave launched from the left, partially obstructed by the pattern's bias field. Magnetostatic potential on the right is produced by the interference of contributions coming from the multiple "open areas" where the pattern's magnetization matches the predominant direction. **(b)** Out-of-plane bias field component produced in the propagation layer directly underneath the recorded pattern. Separation layer thickness is 50nm. The in-plane component's peak value is about 1% of that of the out-of-plane field. There is a flux closure perpendicular component oriented along $H_0$ whose peak value is about +25 Oe.

Consider the magnetization recording pattern shown in (Fig. 4.a) and the respective magnetic bias field it generates (Fig. 4.b). The average out-of-plane component directly underneath the pattern is sufficiently large (-475 Oe) to force the magnetic refractive index into imaginary values for almost any operating frequency. Applying the out-of-plane uniform external field $H_0 = 2900$ Oe ensures the equilibrium magnetization in the propagation layer remains oriented along the z axis. In the first order approximation the segments between the recorded patterns (Fig.4.a) are considered as "open" to spin wave propagation and magnetic potential at an arbitrary point $r$ is given by Kirchhoff-Fresnel integral over such "open" segments:

$$\varphi(\mathbf{r}) \sim \int_S (1 + \cos\theta) H_0^1(k|\mathbf{r} - \acute{\mathbf{r}}|) d\acute{\mathbf{r}} \qquad (18).$$

For the in-plane configuration the wavevector $k$ is angle dependent (Eq.14), but for the out-of-plane case we can assume $k = const = k_0$. Dynamic magnetization components are produced from the magnetostatic potential (Eq.18) by applying (Eq.1) to $\mathbf{h} = \nabla\varphi$:

$$4\pi m_x = \varphi(\mathbf{r})\left((1+\chi)k_x - i\kappa k_y\right), \qquad 4\pi m_y = \varphi(\mathbf{r})\left((1+\chi)k_y + i\kappa k_x\right) \qquad (19).$$

Particular magnetization pattern shown in (Fig. 4) has been optimized using (Eq.18) so that its bias field acts as a Fresnel lens (Fig.5.a); the corresponding distribution of spin wave amplitudes is also computed micromagnetically (Fig.5.b). Here and in the following cases (Figs. 5-7) the source of plane waves is a simple application of a uniform in-plane RF field along the line x=-2000nm.

The difference between the analytical and the micromagnetic results (Fig. 5) is attributable to the fact that the "open" areas still experience some bias field (between +20 and -50 Oe) which perturbs the value of the magnetic refractive index.

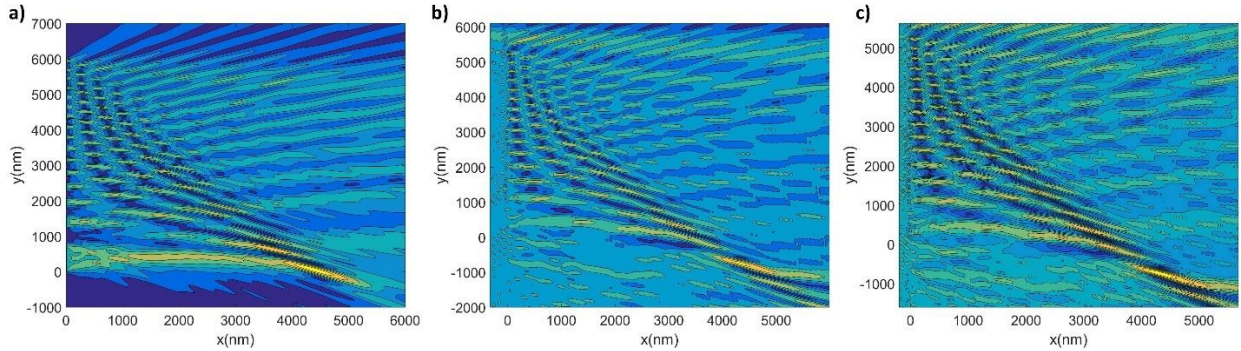

**Figure 5**. Magnetic oscillation amplitudes for the recorded pattern in (Fig.4), computed **(a)** by using (Eq.18), **(b)** micromagnetically **(c)** by solving (Eq.16) via FDTD method. $\omega = 4.9$ GHz and $H_0 = 2900$ Oe.

Much better match between micromagnetics and the wave equation based approach is obtained by solving (Eq.16) via Finite Difference Time Domain (FDTD) technique[26]. As mentioned before, the refractive index is calculated using (Eqs.8-9) except for the vicinity of $\omega_0(\omega_M + \omega_0)$ where numerically fitted values are employed (Fig.2) instead.

There is a question of what exactly one accepts as the value of $\Delta H$ in a system where the bias field has all three spatial components, more so if the bias field value is sufficient to perturb the equilibrium magnetization configuration. General approach would be to first numerically solve for the equilibrium magnetization configuration in the propagation layer in the presence of the actual bias field. Then at every point one calculates the effective $\Delta H$ in (Eq.8) as the difference between the actual local static magnetic field and the uniform field which would exist in a homogeneously saturated sample. This gives rise to an exchange contribution proportional to $\left(\frac{\partial^2 M_x}{\partial x^2} + \frac{\partial^2 M_y}{\partial y^2} + \frac{\partial^2 M_z}{\partial z^2}\right)$, which for a non-uniform magnetization is likely to push the refractive index into the imaginary range. Despite the seeming complexity the technique allows for (Eq.16) to be applied even to systems with a non-uniform topology, such as physical gaps.

On the other hand, the most simplistic yet effective interpretation (used to compute the results in Fig.5.c) assigns to $\Delta H$ the amplitude of the actual bias field irrespective of its direction and without computing the equilibrium magnetization configuration. We have purposefully designed the magnetization pattern in such a way that the out-of-plane component of the field dominates as the flux closure occurs at large enough distance and the peak out-of-plane component is small; in the areas where the in-plane $\Delta H$ is large enough to affect the equilibrium it produces an imaginary $n$ which has limited impact on the overall propagation.

For a substantially complex system there is no single best method to quantitatively estimate the resulting error, i.e. divergence with respect to the micromagnetics. One possibility is to calculate the peak amplitude's location (focal point) in (Fig.5) for a number of different patterns. Analytics (Eq.18) incurs significant relative positioning error, but with FDTD the peak discrepancy along either coordinate axis drops to 9%, with average discrepancy about 2.5%. Noted, in terms of computing speed the wave equation solver is about thousand times faster since the local magnetic field is evaluated only once when the refractive index distribution is being determined (Eqs. 8-9) and there is no need for zero padding and other adjustments. The source of RF field can be introduced in many ways – directly as the incoming plane wave (Eq.22, Fig.6.d), by matching the magnetostatic potential to that of a wire field[10] or by simply setting the value of the magnetostatic potential to a constant along some area which thus serves as the source of RF excitation.

Important advantage of micromagnetics is its ability to handle non-linear problems. It can be shown that for a number of practical cases the wave equation can be adopted to include nonlinearities by making the refractive index dependent on the amplitude of magnetostatic potential $|\varphi|$, for example:

$$n = n(|\varphi| = 0) + \acute{n}|\varphi|^2 + i\acute{n}|\varphi|^2 \tag{20}$$

Instead of the "damping" term proportional to $\acute{n}$ it is also possible to use the term responsible for the generation of different harmonics, converting (Eq.12) into a system of coupled equations for magnetostatic potentials at different frequencies.

Fundamental question is whether the sign of $\acute{n}$ is positive or negative, i.e. whether the high amplitude excitation causes focusing (as common with optical systems) or defocusing? The main source of the refractive index change is the effective saturation magnetization's reduction; which by itself per (Eq.8) reduces the value of $\chi$ and increases n; however in the out-of-plane configuration it also significantly reduces the local demagnetization field, which reduces n. As a result both nonlinear focusing and defocusing can be demonstrated in modeling, depending on the system's configuration and operating parameters.

**4. Scattering.**

Known problem of considerable importance in the field of spin wave dynamics is interaction between a spin wave and various inhomogenuities whose value affects the value of $\chi$, such as the bias field, the saturation magnetization or the anisotropy. Resulting impact is usually considered in terms of the resonance spectrum broadening, i.e. increased effective damping[27]. Most common approach relies on using the perturbation theory to solve a linearized Landau-Lifshitz equation in Fourier space. It performs well with multiple inhomogenuities whose properties are associated with localized distributions in Fourier space, but has limited ability to account for a few, spatially localized inhomogenuities within a large system.

Wave equation (Eq.12) allows one to borrow the scattering methods previously developed in optics and quantum mechanics. Overall solution is decomposed into a sum of a known incident wave $\varphi_I = e^{i(k_{0x}x + k_{0y}x)}$ and an unknown scattering portion $\varphi_S$:

$$\frac{1}{n_y^2}\frac{\partial^2(\varphi_I+\varphi_S)}{\partial y^2} + \frac{1}{n_x^2}\frac{\partial^2(\varphi_I+\varphi_S)}{\partial x^2} + k_0^2(\varphi_I + \varphi_S) = 0 \tag{21}$$

Expressing the refractive indices $n_{0x}, n_{0y}$ as a baseline value plus a perturbation, such as $\frac{1}{n_y^2} = \left(\frac{1}{n_y^2} - \frac{1}{n_{0y}^2}\right) + \frac{1}{n_{0y}^2}$, selecting the incident wavevector $k_0$ as the "baseline" propagation wavevector corresponding to $n_{0x}$ and $n_{0y}$, so that $\frac{1}{n_{0y}^2}\frac{\partial^2 \varphi_I}{\partial y^2} + \frac{1}{n_{0x}^2}\frac{\partial^2 \varphi_I}{\partial x^2} + k_0^2 \varphi_I = 0$, (Eq.21) becomes:

$$-e^{i(k_{0x}x+k_{0y}y)}\left(\left(\frac{1}{n_y^2} - \frac{1}{n_{0y}^2}\right)k_{0y}^2 + \left(\frac{1}{n_x^2} - \frac{1}{n_{0x}^2}\right)k_{0x}^2\right) + \frac{1}{n_y^2}\frac{\partial^2 \varphi_S}{\partial y^2} + \frac{1}{n_x^2}\frac{\partial^2 \varphi_S}{\partial x^2} = -k_0^2 \varphi_S \qquad (22).$$

If needed, in a manner similar to (Eq.15) one can use phenomenological substitution $k_0^2 \varphi_S = -\frac{1}{c_0^2}\frac{\partial^2 \varphi(x,y,t)}{\partial t^2}$ or SVEA approximation. (Eq.22) can also be converted into an integral form using Green's function formalism:

$$\varphi_S(\boldsymbol{r}) = \int_S e^{i(k_{0x}\acute{x}+k_{0y}\acute{y})}\left(\left(\frac{1}{n_y^2} - \frac{1}{n_{0y}^2}\right)k_{0y}^2 + \left(\frac{1}{n_x^2} - \frac{1}{n_{0x}^2}\right)k_{0x}^2\right)H_0^1(k|\boldsymbol{r}-\acute{\boldsymbol{r}}|)d\acute{\boldsymbol{r}} \qquad (23).$$

Born approximation assumes $k = k_0$, which is applicable when second order processes involving consequential scattering on multiple inhomogenuities can be neglected, i.e. areas where magnetic refractive index is perturbed are small in size and well separated from each other, or the amplitude of perturbation is small.

In the out-of-plane configuration the refractive indices are the same along both x and y axis, negating the angle dependence in (Eq.23). Using Born approximation and considering a single inhomogeneity whose size is small compared to the wavelength, i.e. in Rayleigh scattering approximation:

$$\varphi_S(\boldsymbol{r}) \sim k_0^2 \frac{\chi-\chi_0}{1+\chi_0} H_0^1(k_0|\boldsymbol{r}|) \qquad (24),$$

which in the long wavelength limit behaves approximately as $\varphi_S(\boldsymbol{r}) \sim \frac{1}{\lambda^2}$.

When the size of the perturbed area is comparable to the wavelength, the treatment is analogous to Mie scattering in optics. One can solve analytically the case when magnetic refractive index is uniformly perturbed within a circle area by writing in cylindrical coordinates $\varphi_S$ as the sum of outgoing waves outside the circle and standing waves inside the circle, connecting the two solutions via boundary conditions and obtaining an analytical, though overly reliant on special functions, expression for the $\varphi_S$ as a function of the magnetic refractive index perturbation, its radius and the wavelength of the incoming spin waves. Alternatively one can use micromagnetics or solve (Eqs. 15-16) via FDTD (Fig.6).

However, majority of optical methods and designs assume the difference in the refractive index is due to presence of a physical object, i.e. an inhomogeneity which is characterized by a specific size, specific boundaries and made from a material with a homogeneous refractive index. Producing the bias field with a large enough gradient approximating discontinuous "optics-like" boundaries is impractical. Also, simply using a recorded pattern with the shape identical to that of the desired field distribution produces additional flux closure fields which significantly affect the performance except for a few, most simple cases.

However, it is possible to calculate the optimal recording pattern corresponding to a specific magnetic refractive index distribution.

First, the desired distribution of the refractive index $n$ is converted into that of the bias field $\Delta H$ (Fig. 2). Second, a Gaussian blur or another smoothing function is applied at the boundaries, ensuring the maximum magnetic field gradient does not exceed some threshold value, for example 30 Oe/nm, and therefore the peak required average magnetic moment in the recorded pattern (Eq.25) does not exceed the saturation magnetization value.

To derive the corresponding distribution of the magnetization in the hard layer $\boldsymbol{M}(x,y)$ we note that in micromagnetic modeling the dipole-dipole field is calculated by using the formula $\Delta \boldsymbol{H}(x,y) = \int \boldsymbol{A}(x-\acute{x}, y-\acute{y})\boldsymbol{M}(\acute{x},\acute{y})d\acute{x}d\acute{y}$, where $\boldsymbol{A}$ is the demagnetization tensor for the given separation between the layers. Its value is typically computed numerically[16]. In the Fourier space this becomes $\boldsymbol{M}(\boldsymbol{k}) = \Delta \boldsymbol{H}(\boldsymbol{k})/\boldsymbol{A}(\boldsymbol{k})$, and accordingly:

$$\boldsymbol{M}(x,y) = \mathcal{F}\left(\frac{\Delta H(\boldsymbol{k})}{A(\boldsymbol{k})}\right) \qquad (25),$$

which thus determines the magnetization pattern for an arbitrary distribution of the refractive index.

So far in the recorded pattern (Fig. 4) each portion was magnetized in either "up" or "down" direction, however (Eq.25) demands that instead the magnetization behaves as a continuous function of the spatial coordinates. One way to approximate a continuous function $\boldsymbol{M}(x,y)$ via a granular material where each individual grain has only two ("up" and "down") available states is to use a probabilistic recording. For each grain the average value of the magnetization $\boldsymbol{M}(x,y)$ determines the probability of its magnetization to be in either "up" or "down" state. Then, by using Heat Assisted Magnetic Recording process[17], in which the hard media is first heated above the Curie temperature and then its magnetization is set via writer's magnetic field, and decreasing either the magnetic field or the maximum temperature results one can initiate a controllable probabilistic switching, allowing one to record the patterns such as (Fig. 6.a) or (Fig. 7.a) with the area-averaged value of the magnetization roughly matching $\boldsymbol{M}(x,y)$.

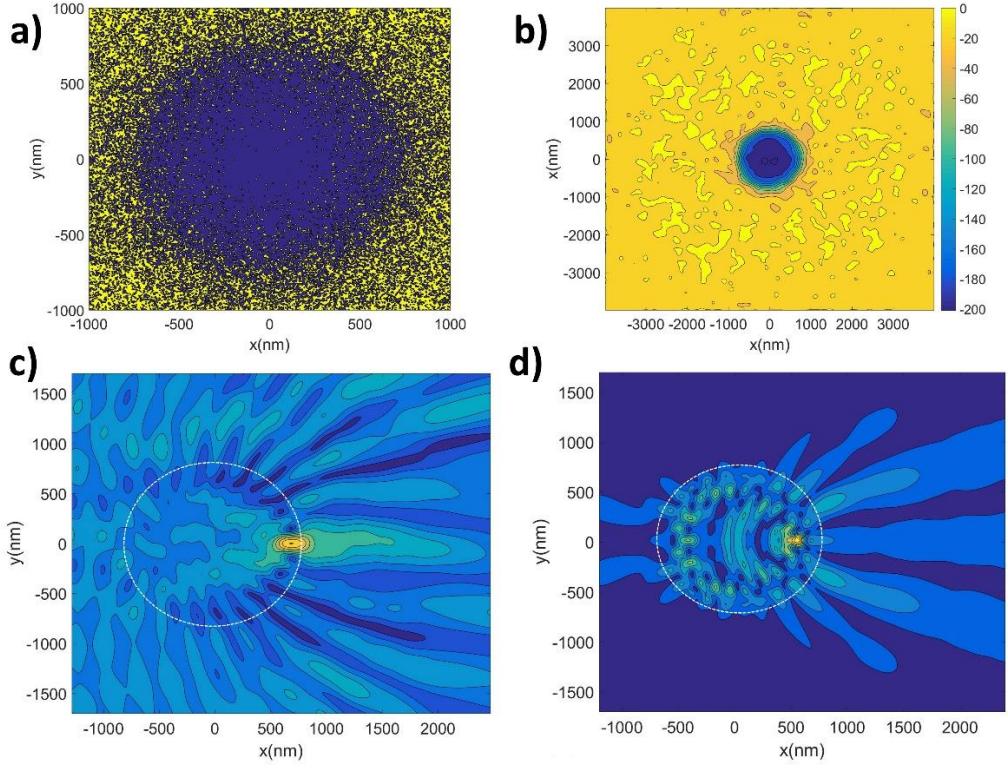

**Figure 6. (a)** Grains switched along (yellow) and against (blue) the applied field $H_0 = 2900$ Oe, **(b)** the resulting bias field in the propagation layer, 100nm below the bias layer **(c)** micromagnetic solution corresponding to Mie resonance within the propagation layer where the bias field is being applied, $\omega = 3.6$ GHz, **(d)** FDTD solution of (Eq.22) for the same operating frequency, using the uniform perturbation bias field $\Delta H = -195$ Oe applied within a white dashed circle 1425nm in diameter.

By using such technique it is possible to produce the localized bias magnetic field (Fig. 6.b.) with mostly uniform properties and observe Mie resonance, i.e. coupling between the magnetic modes localized within the bias field area (Fig.6.c-d) and a plane wave launched from the left. Such localized modes can serve as RF sources as little as 50nm in diameter (Fig. 6.c); in addition the bias field can be used to control the nonlinear terms, suppressing them to such an extent that with the material such as permalloy the localized excitation of this type can reach very high amplitudes, producing RF fields with more than thousand Oe amplitude at 50nm distance away from the propagation layer.

### 4. Refraction and geometric optics.

Consider the case where the wavelength is small compared to the characteristic scale at which $\chi$ varies. The governing equation can be obtained from (Eq.12) using Eikonal approach, i.e. assuming the form $\varphi = \varphi_0 e^{ik_0 \mathcal{L}(x,y)}$ and neglecting[18] all terms except those proportional to $k_0^2$:

$$\frac{1}{n_x^2}\left(\frac{\partial \mathcal{L}}{\partial x}\right)^2 + \frac{1}{n_y^2}\left(\frac{\partial \mathcal{L}}{\partial y}\right)^2 = 1 \qquad (25),$$

which is identical to Eikonal equation for optics[18]. For the out-of-plane configuration the expression is simplified since $n_x = n_y$. Geometric optics in its entirety can be derived from Eikonal equation[18] and therefore in accordance to (Eq.25) can be used in magnetics as well.

This allows for an alternative technique compared to the previously proposed[23,24,25] to create lenses and other devices whose behavior can be modeled via the raytracing (Fig. 7) by using the described above method of converting a refractive index distribution into a recorded pattern. There are obvious distortions caused by the probabilistic recording process (Fig. 7.b); as with other devices whose design is adopted from optics, performance can be improved compared to (Fig. 7) if the distribution of the refractive index is apriori optimized as a continuous variable - something both possible and practical in "spin optics" but difficult to accomplish in actual optics where refractive index values are determined by the material used.

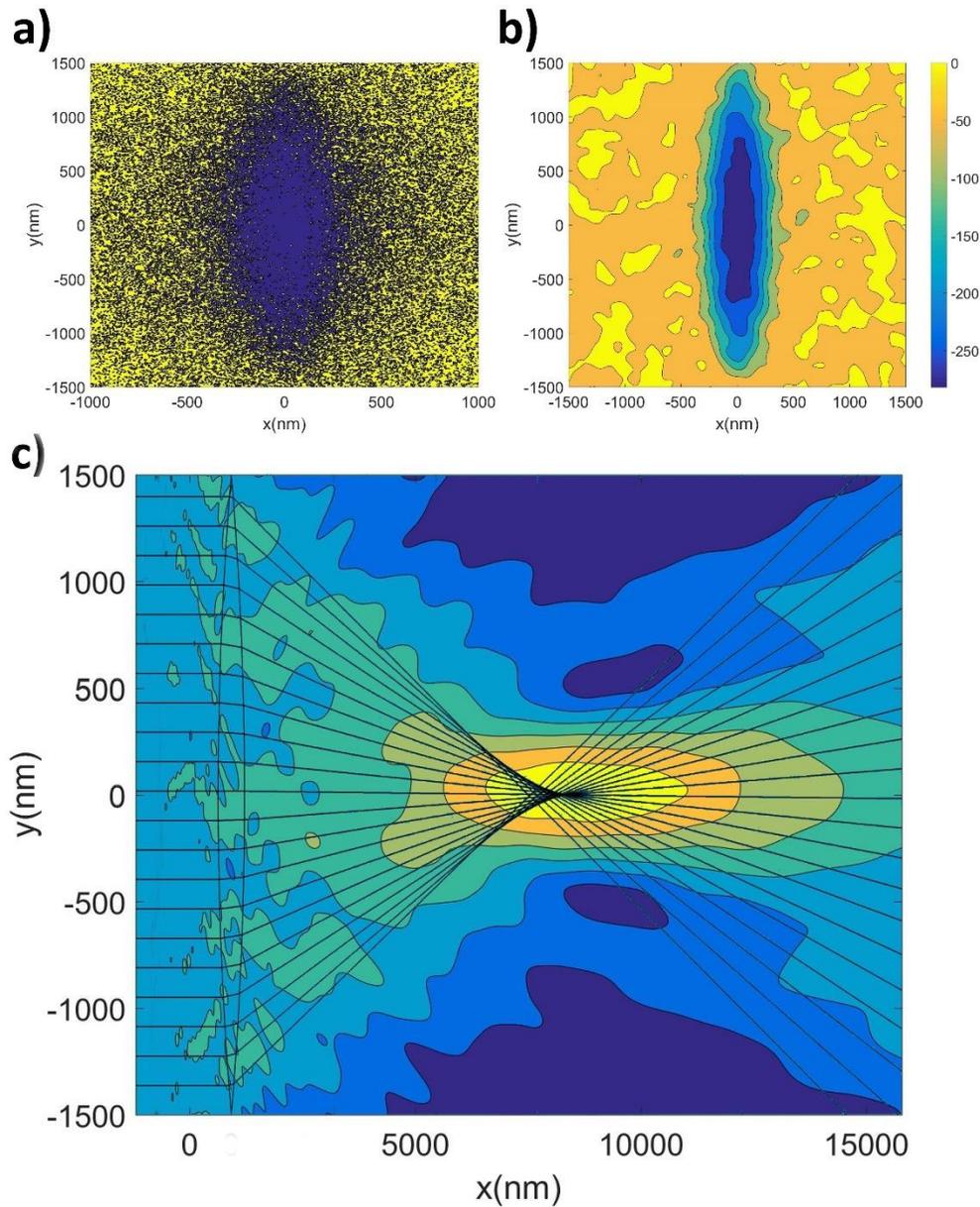

**Figure 7. (a)** Magnetization profile recorded in the hard layer, blue polarity denotes the direction opposite with respect to the applied field $H_0 = 2900$ Oe. Desired curvature radii $R_{1,2} = 4.5$ micrometers, lens height – 3 micrometers. **(b)** Resulting bias field profile in the propagation layer. **(c)** Micromagnetically computed oscillation amplitudes with superimposed solution of the raytracing model. Effective refractive index $n = 1.35$. $\omega = 2.4$ GHz.

When it comes to the minimal achievable focusing spot, both refractive lens (Fig. 7) and Fresnel lens (Fig. 5) perform worse compared to Mie resonance, which is consistent with optical solutions where similar localized modes (Fig. 6) and the related near field phenomena allow for the tighter energy localization.

## 5. Conclusions.

Adopting the wave equation to describe propagation of spin waves allows one to borrow from optics a number of valuable methods and devices. As demonstrated in the current article the former includes FDTD solvers for the wave equation, Green's function based formalism, Born approximation, Rayleigh and Mie scattering, raytracing. In terms of actual devices, numerous examples of optics-like behavior have been previously demonstrated in both modeling and experiment, but using the bias field created by a magnetization pattern recorded onto a neighboring hard layer allows for a generic, deterministic algorithm to convert a near arbitrary two dimensional optical instrument into its spin wave analogue.

Besides YIG there are multiple other propagation layer materials worth considering. FeCoB or Permalloy allow for larger saturation magnetization which can be useful for both RF generation and reception. Recently, organic based low damping compounds[28] have demonstrated significant progress; their advantages include simplicity of deposition coupled with the ability to produce high performing film with little dependence on the roughness of the material located underneath.